\documentclass[aps,superscriptaddress,twocolumn]{revtex4}

\usepackage{amsmath,amssymb}
\usepackage{graphicx}
\usepackage{textcomp}
\usepackage{color}

\newcommand{\eq}[1]{Eq.~(\ref{#1})}
\newcommand{\fig}[1]{Fig.~\ref{#1}}

\begin{document}

\title{Unifying autocatalytic and zeroth order \\branching models for growing actin networks}
\author{Julian Weichsel}
\email{weichsel@berkeley.edu}
\affiliation{Bioquant and Institute for Theoretical Physics, University of Heidelberg, Germany}
\affiliation{Department of Chemistry, University of California at Berkeley, United States}
\author{Krzysztof Baczynski}
\affiliation{Bioquant and Institute for Theoretical Physics, University of Heidelberg, Germany}
\author{Ulrich S. Schwarz}
\email{ulrich.schwarz@bioquant.uni-heidelberg.de}
\affiliation{Bioquant and Institute for Theoretical Physics, University of Heidelberg, Germany}

\date{\today}

\begin{abstract}
The directed polymerization of actin networks is an essential element
of many biological processes, including cell migration. Different
theoretical models considering the interplay between the underlying
processes of polymerization, capping and branching have resulted in
conflicting predictions. One of the main reasons for this discrepancy
is the assumption of a branching reaction that is either first order
(autocatalytic) or zeroth order in the number of existing
filaments. Here we introduce a unifying framework from which the two
established scenarios emerge as limiting cases for low and high
filament number. A smooth transition between the two cases is found
at intermediate conditions. We also derive a threshold for the capping
rate, above which autocatalytic growth is predicted at sufficiently
low filament number. Below the threshold, zeroth order
characteristics are predicted to dominate the dynamics of the network
for all accessible filament numbers. Together, this allows cells to
grow stable actin networks over a large range of different conditions.
\end{abstract}

\maketitle

\begin{figure}[b]
\includegraphics[width=\columnwidth]{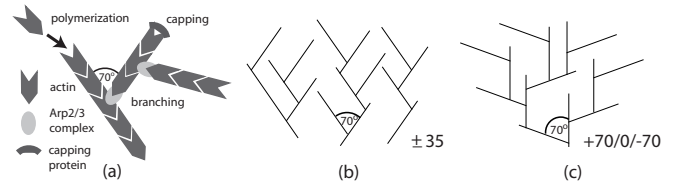}
\caption{(a) Interplay of polymerization, capping and branching at
the leading edge of an actin network growing towards the top.
(b) A $\pm35$ pattern is usually associated
with dendritic actin networks. (c) However, theoretical
and experimental evidence also exists for a
$+70/0/\!\!-\!\!70$ pattern.}
\label{cartoon}
\end{figure}

In many situations of high biological relevance, including the
migration of animal cells and the propulsion of specific intracellular
pathogens, motility results from the directed polymerization of a
dendritic actin filament network \cite{carlier2010}. The organization
of the growing network is determined mainly at the leading edge, where
a small number of proteins regulates the interplay between three
fundamental processes. The driving force for propulsion is
polymerization of actin filaments from globular actin monomers. This
is limited by capping proteins, which bind to the filament ends and
prevent further polymerization. New filaments nucleate by branching
off from mother filaments \cite{pollard2007}. Although the biochemical
details of this process are not yet completely understood, it is widely
accepted that the branching complex Arp2/3 is activated by nucleation
promoting factors (NPFs) like WASP and SCAR/WAVE proteins
\cite{beltzner2008,xu2012}. When an activated Arp2/3-complex is bound
to the side of an existing actin filament, a daughter filament starts
to grow at a characteristic angle around $70^\circ$ relative to the
mother filament (compare \fig{cartoon}a). At the same time, the
branch point moves away from the leading edge because of the
on-going polymerization of actin filaments.

Due to the high biological relevance and universal nature of the
underlying processes, many theoretical models have been suggested to
describe the characteristic features of growing actin networks
\cite{mogilner2009}.  However, in many cases contradictory predictions
have been obtained, in particular regarding experimentally observed
force-velocity relations
\cite{wiesner2003,mcgrath2003,marcy2004,parekh2005,prass2006,heinemann2011,zimmermann2012}
and the filament orientation distribution of the network
\cite{maly2001,verkhovsky2003,schaub2007,koestler2008,weichsel2012}. Interestingly,
many of these contradictions are a direct consequence of two different
choices for the order of the branching reaction. In
\textit{autocatalytic models}, the branching rate is assumed to be
proportional to the number of existing filaments in the network,
i.e. it is modeled as a {\it first order} reaction in filament
density, implicitly assuming an unlimited reservoir of activated
Arp2/3 \cite{maly2001,carlsson2003,schaus2007}.  This yields growing
actin networks for which a constant filament density is maintained
only at a unique steady state growth velocity.  Increasing forces
acting against the network reduce the speed of growth only
transiently, as an increasing filament density subsequently lowers the
force per filament back to the stationary level.

In marked contrast to the autocatalytic scenario, another class of
models assumes that branching occurs with a constant rate, i.e. it is
taken to be a {\it zeroth order} reaction in filament density,
corresponding to a limited supply of activated Arp2/3
\cite{carlsson2003,schaub2007,weichsel2010_2}.  Under these
conditions, it has been shown that a continuum of steady state
velocities exists. Moreover, two competing steady state filament
orientation patterns are stable, namely the $\pm35$ and
$+70/0/\!\!-\!\!70$ patterns shown schematically in \fig{cartoon}b and
c, respectively.  Transitions between these two fundamentally
different network architectures can be triggered by changes in network
growth velocity \cite{maly2001,weichsel2010_2}. Indeed similar structural
transitions have been demonstrated recently in electron microscopy data
of the lamellipodium of keratocytes, indicating their physiological relevance
\cite{koestler2008,weichsel2012}. In this Letter, we will show that
the two contradictory model scenarios of autocatalytic and zeroth
order branching can be unified within a general theoretical framework
that reconciles some of the seemingly contradictory observations and
predictions.

\emph{Arp2/3 activation model.} We first introduce a kinetic model for
filament branching, based on a likely scenario for Arp2/3 activation
\cite{beltzner2008,ti2011,xu2012}. Motivated by the dimensions of the
lamellipodium for cells migrating on a flat substrate, we consider a
two-dimensional situation in which the network moves away from the
leading edge with a well defined retrograde velocity $v_{\rm nw}$. All
reactions are assumed to occur in a small reaction zone extending from
the leading edge over a nanometer-scale distance $d_{\rm br}$. We
consider a system of two variables: $A$ is the concentration of Arp2/3
that is bound to the filaments, but did not lead to a daughter branch
yet. $P$ is the concentration of NPFs which is available to activate
bound Arp2/3 complexes to nucleate a daughter branch. The kinetic
equations are
\begin{equation}
\begin{array}{lcl}
\frac{d A}{dt} & = & k_{+} N_{\rm fil} - \left(k_{-} + \frac{v_{\rm nw}}{d_{\rm br}}\right)  A - \tilde{k}_{\rm b}   A \, P\ , \\ 
\frac{d P}{dt} & = & - \tilde{k}_{\rm b} A \, P + k_{\rm act} \left( P_{\rm 0} - P \right)\ .
\end{array}
\label{arp_activation_dgl} 
\end{equation}
$A$ increases as more complexes bind to the filaments with rate
$k_{+}$ and decreases due to dissociation (rate $k_{-}$), outgrowth
(rate $v_{\rm nw} / d_{\rm br}$) and branching (rate $\tilde{k}_{\rm
  b}$). The last step also decreases available $P$, as NPFs that
activate Arp2/3 are occupied for additional interactions with other
Arp2/3 complexes at the same time until they become available again at
rate $k_{\rm act}$. $P_{\rm 0}$ is the total concentration of NPFs and
$N_{\rm fil}$ is the number of actin filaments (because $N_{\rm fil}$ will
be a central quantity of interest below, for our purpose it is convenient
to consider the number of filaments in a reaction volume of finite
lateral size rather than their concentration).

\begin{figure}[t]
\includegraphics[width=0.9\columnwidth]{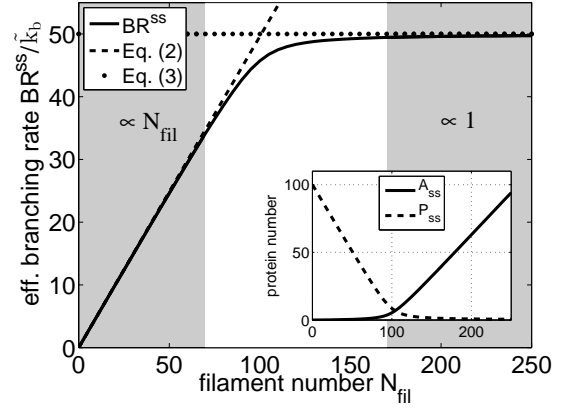}
\caption{Effective branching rate $\mathrm{BR}^{\rm ss}$ versus
  filament number $N_{\rm fil}$ in steady state.  At low filament
  number the branching reaction is linear (autocatalytic) as given in
  \eq{BR_lowN} (dashed line), while at high filament number, a zeroth
  order branching reaction is observed with a constant rate given by
  \eq{BR_highN} (dotted line). The gray backgrounds mark the
  first and zeroth order regimes. The inset shows the
  corresponding steady state concentrations of filament bound Arp2/3
  ($A_{\rm ss}$, solid line) and available NPFs ($P_{\rm ss}$, dashed
  line).}
\label{fig_BR_vs_Nfil}
\end{figure}

In steady state, \eq{arp_activation_dgl} defines an effective rate of
branching as $\mathrm{BR}^{\rm ss} \equiv \tilde{k}_{\rm b} A_{\rm ss}
P_{\rm ss}$. This rate is a function of filament number $N_{\rm fil}$
as plotted in \fig{fig_BR_vs_Nfil} for a typical set of parameters.
At sufficiently small filament number $N_{\rm fil}$, the effective
branching rate is approximately first order in $N_{\rm fil}$ and hence
the coefficient of its linear expansion defines an autocatalytic
branching rate constant $k_{\rm b}^{\rm ac}$:
\begin{equation}
\mathrm{BR}^{\rm ss}_{0}  = 
\frac{P_{0} d_{\rm br} k_{+} \tilde{k}_{\rm b}}{P_{\rm 0} d_{\rm br} \tilde{k}_{\rm b} + d_{\rm br} k_{-} + v_{\rm nw}}
N_{\rm fil} \equiv k_{\rm b}^{\rm ac} N_{\rm fil}\ .
\label{BR_lowN}
\end{equation}
In the limit of large filament number $N_{\rm fil}$, $\mathrm{BR}^{\rm ss}$ saturates at a constant rate
as assumed in zeroth order branching models:
\begin{equation}
\mathrm{BR}^{\rm ss}_{\infty} = P_{\rm 0} k_{\rm act}\ .
\label{BR_highN}
\end{equation}
Thus the reaction smoothly changes from 
first to zeroth order as the filament number $N_{\rm fil}$ increases.

\emph{Actin growth model.} We next analyze the effect of the order of
the branching reaction on the steady growth states of actin
networks. To this end, we extend a deterministic rate
equation model that has been used before to describe both
autocatalytic as well as zeroth order branching actin networks
\cite{maly2001,carlsson2003,weichsel2010_2}.  The generic results
reported here can be confirmed in computer simulations based on
individual filaments and stochastic reactions \cite{si}. We consider an ensemble of
filaments located in the same reaction zone of width $d_{\rm br}$ as
introduced above. Our central quantity is the distribution function
$N(\theta,t)$ for the number of uncapped filaments orientated
at time $t$ at an angle $\theta$ with respect to the normal of the
leading edge, which evolves in time as
\begin{equation}
\begin{array}{rcl}
\frac{dN(\theta,t)}{dt} & = & - k_{\rm c} N(\theta,t) - k_{\rm gr}^{\theta}(v_{\rm nw}) N(\theta,t) \\
& & + k_{\rm b}  \frac{\int\limits_{-\pi}^{+\pi} \mathcal{W}(\theta,\theta') N(\theta',t)
\,\mbox{d}\theta'}{\left( \int\limits_{-\pi}^{+\pi}\int\limits_{-\pi}^{+\pi}
\mathcal{W}(\theta,\theta') N(\theta',t) \,\mbox{d}\theta'\,\mbox{d}\theta \right)^{1 - \mu}}\ .
\end{array}
\label{integr_dgl_fiber_orient} 
\end{equation}
Here the three terms on the right introduce capping, outgrowth from
the reaction zone and branching, respectively. While capping is simply
a first order process with constant rate, independent of filament
orientation $\theta$, for outgrowth we have to distinguish two
cases. For $\vert\theta\vert \leq \arccos (v_{\rm nw} / v_{\rm fil})$,
single filaments growing with velocity $v_{\rm fil}$ can keep up with
the leading edge and thus $k_{\rm gr}^{\theta}(v_{\rm nw}) = 0$. If
the orientation angle exceeds the threshold, filaments grow too slowly
and leave the reaction region with rate $k_{\rm gr}^{\theta}(v_{\rm
  nw}) = (v_{\rm nw}-v_{\rm fil} \cos{\theta}) / d_{\rm br}$.

In the branching term,
$\mathcal{W}(\theta,\theta')=\mathcal{W}(|\theta-\theta'|)$ is a
distribution function of the relative branching angle between mother
and daughter filaments. Motivated by experimental observations, we
approximate this function by the sum of two Gaussians centered around
$\pm 70^\circ$ and each with standard deviation $5^\circ$. This
corresponds well to the experimentally reported range of branching
angles between $67^\circ$ and $77^\circ$, that have been measured both
using purified proteins \cite{mullins1998,blanchoin2000} and in
different cell lines \cite{svitkina1999,vinzenz2012}. The exact value
of the branching angle, however, is irrelevant for our results.

The normalization of the branching term in
\eq{integr_dgl_fiber_orient} is appropriate to directly implement a
specific reaction order $\mu$ in the actin growth model. In order to
couple the actin growth model to the Arp2/3 activation model, below we
will use numerical calculations with $\mu=0$ and couple
\eq{arp_activation_dgl} and \eq{integr_dgl_fiber_orient} via the
filament number dependent branching rate $k_{\rm b}(N_{\rm
  fil})=\mathrm{BR}^{\rm ss}$ derived above, with $N_{\rm fil} = \int
N(\theta,t) d\theta$. For analytical progress and deeper insight,
however, it is instructive to first analyze the steady states of the
actin growth model with constant parameters $k_{\rm b}$ and $\mu$. Indeed, it
can be shown within the linear stability analysis employed below that
both procedures are equivalent \cite{si}.

\emph{Steady state analysis.} The actin growth model
\eq{integr_dgl_fiber_orient} contains only four relevant parameters,
the rates $k_{\rm c}$ and $k_{\rm b}$ for capping and branching,
respectively, the network growth velocity $v_{\rm nw}$ and the order
of the branching reaction $\mu$. By integrating the reaction model
\eq{integr_dgl_fiber_orient} over $35^\circ$ sized angle bins and
neglecting contributions from filaments growing in directions $>
87.5^\circ$, we obtain three simplified coupled equations for
the evolution of $N_{0^\circ}$, $N_{\pm35^\circ}$ and
$N_{\pm70^\circ}$, which can be analyzed analytically. As an
alternative which does not require any additional assumptions, we
determine the stable regimes of network growth numerically, by
propagating a finely discretized version of the equation until a
steady state is reached.

\begin{figure}[t]
\includegraphics[width=1\columnwidth]{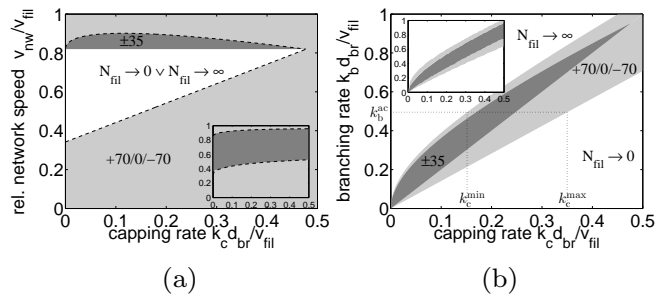}
\caption{Phase diagram predicted from linear stability analysis of the
  analytical model and by numerically solving the actin growth
  model (insets) \cite{si}. (a)
  Projection onto $(k_{\rm c},v_{\rm nw})$-plane. The dashed contour
  indicates identical transitions between $+70/0/\!\!-\!\!70$ and
  $\pm35$ patterns for all $0 \le \mu \le 1$. For $\mu = 1$, an
  additional constraint restricts the stable parameter space to the
  gray shaded regions. (b) Projection onto $(k_{\rm c},k_{\rm
    b})$-plane for $\mu = 1$ (autocatalytic growth). Only a subset of
  the parameter values results in stable steady state solutions.}
\label{fig_phase_diagram}
\end{figure}

In the analytical approach, there exist exactly two physically meaningful
steady state solutions, $N^{\rm ss70}$ and $N^{\rm ss35}$, given by
\begin{equation}
\begin{array}{rcl}
N_{0^\circ}^{\rm ss70} & = & \frac{ - k_{\rm c} - k_{\rm gr}^{70^\circ} + \sqrt{2 k_{\rm c}\left(k_{\rm c}
+ k_{\rm gr}^{70^\circ} \right)}}{k_{\rm c} - k_{\rm gr}^{70^\circ}} \cdot C_{1}^{1/(1-\mu)}\, \\
N_{\pm35^\circ}^{\rm ss70} & = & 0\, \\
N_{\pm70^\circ}^{\rm ss70} & = & \frac{2 k_{\rm c} - \sqrt{2 k_{\rm c}\left(k_{\rm c}
+ k_{\rm gr}^{70^\circ} \right)}}{k_{\rm c} - k_{\rm gr}^{70^\circ}} \cdot C_{1}^{1/(1-\mu)}\, \\
\end{array}
\label{sol070}
\end{equation} 
and
\begin{equation}
N_{0^\circ}^{\rm ss35} = N_{\pm70^\circ}^{\rm ss35}  =  0,\
N_{\pm35^\circ}^{\rm ss35} = C_{2}^{1/(1-\mu)}
\label{sol35}
\end{equation} 
where
\begin{equation}
C_{1} = k_{\rm b} / \sqrt{2 k_{\rm c} \left( k_{\rm c} + k_{\rm gr}^{70^\circ} \right)},\
C_{2} = k_{\rm b} / (2 k_{\rm c} + 2 k_{\rm gr}^{35^\circ}).\
\label{AB}
\end{equation} 
These two fixed points correspond to the two competing orientation
patterns depicted schematically in \fig{cartoon}c and b, respectively.
Linear stability analysis shows that for $\mu > 1$, both are saddle
points and thus no stable solution exists. In contrast, $\mu \leq 1$
leads to mutually exclusive stability of the two solutions
\cite{si}. \fig{fig_phase_diagram}a shows the regions of stability for
each of the two orientation patterns within the two dimensional
parameter space spanned by $k_{\rm c}$ and $v_{\rm nw}$.  The dashed
contour indicates transitions between a $+70/0/\!\!-\!\!70$ pattern
outside and a $\pm35$ pattern inside. Remarkably, the transition is
independent of $k_{\rm b}$ and $\mu$ and thus all cases with $\mu \leq
1$ show no difference in the locations of the
transitions. The result from the full numerical analysis of
\eq{integr_dgl_fiber_orient} is shown as inset. The main difference
between the analytical and the numerical result is that the stability
of the $\pm35$ pattern vanishes for large $k_{\rm c}$ in the analytical
model, because it disregards contributions from filament orientations
$\gtrsim 90^\circ$. This increases stability of the
$+70/0/\!\!-\!\!70$ pattern, when outgrowth of filaments is negligible
compared to capping.

Until now we have shown, that for all reaction orders of interest ($0
\leq \mu \leq 1$), two orientation patterns compete for stability,
with phase boundaries being independent of the exact value of
$\mu$. Nevertheless the limit $\mu \to 1$ (autocatalytic growth) is
special, because in this case, finite steady state solutions only exist
if additional constraints are satisfied. Due to \eq{sol070}, $N^{\rm
  ss70}$ is finite only when $C_{1}=1$. From \eq{AB} this requires
$k_{\rm gr}^{70^\circ}(v_{\rm nw}) k_{\rm c} + k_{\rm c}^2 = k_{\rm
  b}^2 / 2$. Due to \eq{sol35}, $N^{\rm ss35}$ is finite only when
$C_{2}=1$. From \eq{AB} this requires $k_{\rm gr}^{35^\circ}(v_{\rm
  nw}) + k_{\rm c} = k_{\rm b} / 2$. Therefore, if a stable
steady state solution exists for given values of $k_{\rm c}$ and
$k_{\rm b}$, then for $\mu = 1$ it corresponds to a unique network
growth velocity $v_{\rm nw}$. In this way, the most prominent feature
of autocatalytic growth \cite{carlsson2003} emerges in our unifying
model. Due to these additional conditions, stable solutions are
restricted in parameter space to a lower dimensional manifold, which
in case of the analytical model has a jump discontinuity \cite{si}.  In
\fig{fig_phase_diagram}a we show the projection of this manifold onto
the $(k_{\rm c}, v_{\rm nw})$-plane with bright and dark gray regions
marking the stability regions for the $+70/0/\!\!-\!\!70$ and
$\pm35$ patterns, respectively. As the inset indicates, a jump
discontinuity is not observed in the full numerical treatment.

In \fig{fig_phase_diagram}b, the manifold for $\mu = 1$ is projected
onto the $(k_{\rm c},k_{\rm b})$-plane (in this projection, the jump
discontinuity cannot be seen).  Only a subset of ($k_{\rm b},k_{\rm
  c})$-combinations yields stable autocatalytic growth with finite
filament number $N_{\rm fil}$. This important result is predicted
both by the analytical and the numerical approach (compare inset).

\begin{figure}[t]
\includegraphics[width=0.9\columnwidth]{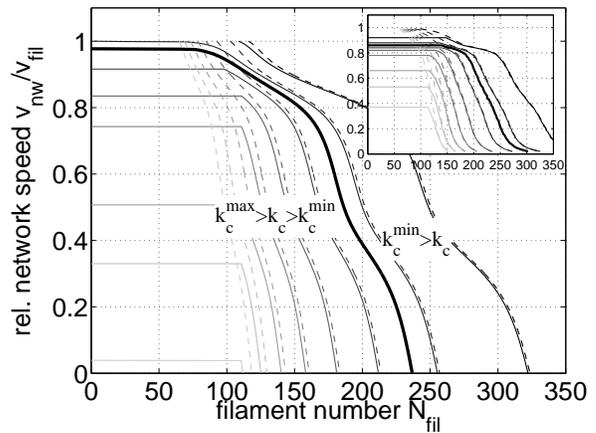}
\caption{Network growth velocity as a function of filament number for
  different capping rates $k_{\rm c}$ as obtained numerically from the
  unified model (solid lines). Darker gray is indicating decreasing
  $k_{\rm c}$. The dashed lines show the results from a zeroth order
  description. For capping rates $k_{\rm c}^{\rm min} \lesssim k_{\rm
    c} \lesssim k_{\rm c}^{\rm max}$, an autocatalytic regime is
  observed at low filament density.  The capping rate $k_{\rm
    c}=k_{\rm c}^{\rm min}$ (thick black solid line) marks the
  transition to pure zeroth order behavior. The inset shows the
  results from stochastic computer simulations.}
\label{fig_Nfil_vs_vnw}
\end{figure}

\emph{Limits of autocatalytic network growth.}  Using the insights
obtained in the preceding sections from the actin growth model
\eq{integr_dgl_fiber_orient} with $\mu$ as a model parameter, we now
combine the Arp2/3 activation model \eq{arp_activation_dgl} and the
actin growth model \eq{integr_dgl_fiber_orient} with $\mu=0$ to arrive
at a unifying theoretical framework for actin network growth with a
branching reaction that is determined by a regulatory process.
\fig{fig_Nfil_vs_vnw} shows our numerical results for network growth
velocity $v_{\rm nw}$ as a function of filament number $N_{\rm fil}$
(solid lines) for various values of the capping rate $k_{\rm c}$. They
agree very well with the results from stochastic computer simulations
shown as inset \cite{si}. At sufficiently low filament number, we
observe an autocatalytic regime where a whole range of values for
$N_{\rm fil}$ corresponds to the same velocity. For larger $N_{\rm
  fil}$, however, the steady state network velocity starts to decrease
similarly to a pure zeroth order description (dashed lines). These
changes include transitions between the two dominant filament
orientation patterns as predicted in \fig{fig_phase_diagram}.

Interestingly, the details of the crossover from first to zeroth order
branching strongly depend on the capping rate. This can be understood
in the analytical model analyzed above. At low filament density,
branching is effectively a first order reaction and thus the
conditions, $C_{1}=1$ and $C_{2}=1$, previously derived from
\eq{sol070}--\eq{AB} for $\mu=1$, need to be satisfied here for stable
growth as well. By inserting the autocatalytic branching rate $k_{\rm
  b}^{\rm ac}$ defined in \eq{BR_lowN} into \eq{AB} and applying the
relevant first order condition, we are able to derive estimates for
minimum and maximum capping rates $k_{\rm c}^{\rm min}$ and $k_{\rm
  c}^{\rm max}$ corresponding to the largest and smallest possible
network velocities, $v_{\rm nw}/v_{\rm fil}=1$ and $v_{\rm nw}/v_{\rm
  fil}=0$, respectively:
\begin{equation}
k_{\rm c}^{\rm min} =  \frac{v_{\rm fil}}{2 d_{\rm br}} \left[ c
+ \sqrt{2 \left( \frac{k_{\rm b}^{\rm ac} d_{\rm br}}{v_{\rm fil}} \right)^2 + c^2} \right],\ 
k_{\rm c}^{\rm max} = \frac{k_{\rm b}^{\rm ac}}{\sqrt{2}}
\label{cappingrate}
\end{equation}
where $c = \cos 70^\circ - 1$. These two threshold values are shown in
\fig{fig_phase_diagram}b as the intersection of $k_{\rm b}^{\rm ac}$
with the boundaries of the stable autocatalytic parameter subset.
Comparison of \eq{cappingrate} with the numerical results from the
full model presented in \fig{fig_Nfil_vs_vnw} shows that our
analytical approach captures the location of this crossover very well
and thus accurately explains the observed behavior.  For increasing
$k_{\rm c}$, the network velocity in the autocatalytic region
decreases until at around $k_{\rm c} \simeq k_{\rm c}^{\rm max}$ the
filament number decays to zero for all accessible network
velocities. For decreasing $k_{\rm c}$, the network growth velocity
reaches its maximal value at $k_{\rm c} \lesssim k_{\rm c}^{\rm min}$
(thick solid line), when the network is not able to balance filament
branching by capping and outgrowth anymore.  In a purely autocatalytic
model, this would lead to a diverging and therefore unphysical
filament number.  Within our unifying framework, the number of
filaments increases only to the point where zeroth order growth
behavior starts to dominate and stabilizes a steady state at finite
filament number. In this regime, the results from the full model
(solid) agree with a zeroth order branching model (dashed).

\emph{Relation to experiments.}  In this Letter, we have developed a
theoretical framework that reconciles conflicting results from two
classes of actin growth models and explains many experimental
observation: an autocatalytic growth regime at low filament density
\cite{wiesner2003,parekh2005}, zeroth order characteristics at high
density \cite{mcgrath2003,marcy2004}, network velocity-dependent
transitions in filament orientation patterns
\cite{koestler2008,weichsel2012} and bistability and hysteresis at
these transitions \cite{parekh2005,weichsel2010_2}. Strikingly our
model naturally avoids the instability which occurs at low capping
rate in the autocatalytic model.

Our model also makes testable predictions that can guide future
experiments.  Using single molecule microscopy either in migrating
cells \cite{millius2012} or in reconstituted assays, the number of
branching events can be directly correlated to filament density, which
can be compared to the effective branching rate as predicted in
\fig{fig_BR_vs_Nfil}. From electron microscopy data, filament
orientations can be extracted and correlated with the growth velocity
as demonstrated in \cite{koestler2008,weichsel2012}. This can be
compared to the unified phase diagram in
\fig{fig_phase_diagram}. Force-velocity relations could be calculated
for our model along the lines of
Refs.~\cite{carlsson2003,weichsel2010_2}, but would require additional
assumptions regarding for example network mechanics, load sharing and
filament-membrane interactions.  However, some general conclusions can
already be drawn at this point and are explained best for the case of
reconstituted actin networks growing against a functionalized AFM
cantilever or bead \cite{marcy2004,parekh2005,chaudhuri2009}. In this
context, our model predicts an autocatalytic (i.e. force-insensitive)
growth velocity for sufficiently low load {\it and} high concentration
of capping protein. In this regime the filament density near the
obstacle is thus expected to grow proportional to the applied
force. When either the concentration of capping protein is reduced
below the threshold $k_{\rm c}^{\rm min}$ (\eq{cappingrate}) or the
load on the network is sufficiently increased, zeroth order behavior
is predicted to take over as is illustrated in \fig{fig_Nfil_vs_vnw}.
If combined with mechanical models like Ref.~\cite{zimmermann2012}, in
the future these kinetic considerations might lead to a complete
understanding of the intriguing physics of growing actin networks.

\begin{acknowledgments}
JW was supported by the research unit for systems biology ViroQuant at
Heidelberg and by the Deutsche Forschungsgemeinschaft (DFG) at
Berkeley (grant no. We 5004/2-1).  USS is member of the cluster of
excellence CellNetworks at Heidelberg University.
\end{acknowledgments}




\begin{thebibliography}{27}
\expandafter\ifx\csname natexlab\endcsname\relax\def\natexlab#1{#1}\fi
\expandafter\ifx\csname bibnamefont\endcsname\relax
  \def\bibnamefont#1{#1}\fi
\expandafter\ifx\csname bibfnamefont\endcsname\relax
  \def\bibfnamefont#1{#1}\fi
\expandafter\ifx\csname citenamefont\endcsname\relax
  \def\citenamefont#1{#1}\fi
\expandafter\ifx\csname url\endcsname\relax
  \def\url#1{\texttt{#1}}\fi
\expandafter\ifx\csname urlprefix\endcsname\relax\def\urlprefix{URL }\fi
\providecommand{\bibinfo}[2]{#2}
\providecommand{\eprint}[2][]{\url{#2}}

\bibitem[{\citenamefont{Carlier}(2010)}]{carlier2010}
\bibinfo{editor}{\bibfnamefont{M.~F.} \bibnamefont{Carlier}}, ed.,
  \emph{\bibinfo{title}{Actin-based Motility}} (\bibinfo{publisher}{Springer
  Netherlands}, \bibinfo{year}{2010}).

\bibitem[{\citenamefont{Pollard}(2007)}]{pollard2007}
\bibinfo{author}{\bibfnamefont{T.~D.} \bibnamefont{Pollard}},
  \bibinfo{journal}{Annual Reviews in Biophysics and Biomolecular Structure}
  \textbf{\bibinfo{volume}{36}}, \bibinfo{pages}{451} (\bibinfo{year}{2007}).

\bibitem[{\citenamefont{Beltzner and Pollard}(2008)}]{beltzner2008}
\bibinfo{author}{\bibfnamefont{C.~C.} \bibnamefont{Beltzner}} \bibnamefont{and}
  \bibinfo{author}{\bibfnamefont{T.~D.} \bibnamefont{Pollard}},
  \bibinfo{journal}{Journal of Biological Chemistry}
  \textbf{\bibinfo{volume}{283}}, \bibinfo{pages}{7135} (\bibinfo{year}{2008}).

\bibitem[{\citenamefont{Xu et~al.}(2012)\citenamefont{Xu, Rouiller, Slaughter,
  Egile, Kim, Unruh, Fan, Pollard, Li, Hanein et~al.}}]{xu2012}
\bibinfo{author}{\bibfnamefont{X.~P.} \bibnamefont{Xu}},
  \bibinfo{author}{\bibfnamefont{I.}~\bibnamefont{Rouiller}},
  \bibinfo{author}{\bibfnamefont{B.~D.} \bibnamefont{Slaughter}},
  \bibinfo{author}{\bibfnamefont{C.}~\bibnamefont{Egile}},
  \bibinfo{author}{\bibfnamefont{E.}~\bibnamefont{Kim}},
  \bibinfo{author}{\bibfnamefont{J.~R.} \bibnamefont{Unruh}},
  \bibinfo{author}{\bibfnamefont{X.}~\bibnamefont{Fan}},
  \bibinfo{author}{\bibfnamefont{T.~D.} \bibnamefont{Pollard}},
  \bibinfo{author}{\bibfnamefont{R.}~\bibnamefont{Li}},
  \bibinfo{author}{\bibfnamefont{D.}~\bibnamefont{Hanein}},
  \bibnamefont{et~al.}, \bibinfo{journal}{The EMBO Journal}
  \textbf{\bibinfo{volume}{31}}, \bibinfo{pages}{236} (\bibinfo{year}{2012}).

\bibitem[{\citenamefont{Mogilner}(2009)}]{mogilner2009}
\bibinfo{author}{\bibfnamefont{A.}~\bibnamefont{Mogilner}},
  \bibinfo{journal}{Journal of Mathematical Biology}
  \textbf{\bibinfo{volume}{58}}, \bibinfo{pages}{105} (\bibinfo{year}{2009}).

\bibitem[{\citenamefont{Wiesner et~al.}(2003)\citenamefont{Wiesner, Helfer,
  Didry, Ducouret, Lafuma, Carlier, and Pantaloni}}]{wiesner2003}
\bibinfo{author}{\bibfnamefont{S.}~\bibnamefont{Wiesner}},
  \bibinfo{author}{\bibfnamefont{E.}~\bibnamefont{Helfer}},
  \bibinfo{author}{\bibfnamefont{D.}~\bibnamefont{Didry}},
  \bibinfo{author}{\bibfnamefont{G.}~\bibnamefont{Ducouret}},
  \bibinfo{author}{\bibfnamefont{F.}~\bibnamefont{Lafuma}},
  \bibinfo{author}{\bibfnamefont{M.~F.} \bibnamefont{Carlier}},
  \bibnamefont{and}
  \bibinfo{author}{\bibfnamefont{D.}~\bibnamefont{Pantaloni}},
  \bibinfo{journal}{Journal of Cell Biology} \textbf{\bibinfo{volume}{160}},
  \bibinfo{pages}{387} (\bibinfo{year}{2003}).

\bibitem[{\citenamefont{McGrath et~al.}(2003)\citenamefont{McGrath,
  Eungdamrong, Fisher, Peng, Mahadevan, Mitchison, and Kuo}}]{mcgrath2003}
\bibinfo{author}{\bibfnamefont{J.~L.} \bibnamefont{McGrath}},
  \bibinfo{author}{\bibfnamefont{N.~J.} \bibnamefont{Eungdamrong}},
  \bibinfo{author}{\bibfnamefont{C.~I.} \bibnamefont{Fisher}},
  \bibinfo{author}{\bibfnamefont{F.}~\bibnamefont{Peng}},
  \bibinfo{author}{\bibfnamefont{L.}~\bibnamefont{Mahadevan}},
  \bibinfo{author}{\bibfnamefont{T.~J.} \bibnamefont{Mitchison}},
  \bibnamefont{and} \bibinfo{author}{\bibfnamefont{S.~C.} \bibnamefont{Kuo}},
  \bibinfo{journal}{Current Biology} \textbf{\bibinfo{volume}{13}},
  \bibinfo{pages}{329} (\bibinfo{year}{2003}).

\bibitem[{\citenamefont{Marcy et~al.}(2004)\citenamefont{Marcy, Prost, Carlier,
  and Sykes}}]{marcy2004}
\bibinfo{author}{\bibfnamefont{Y.}~\bibnamefont{Marcy}},
  \bibinfo{author}{\bibfnamefont{J.}~\bibnamefont{Prost}},
  \bibinfo{author}{\bibfnamefont{M.~F.} \bibnamefont{Carlier}},
  \bibnamefont{and} \bibinfo{author}{\bibfnamefont{C.}~\bibnamefont{Sykes}},
  \bibinfo{journal}{Proceedings of the National Academy of Sciences of the
  United States of America} \textbf{\bibinfo{volume}{101}},
  \bibinfo{pages}{5992} (\bibinfo{year}{2004}).

\bibitem[{\citenamefont{Parekh et~al.}(2005)\citenamefont{Parekh, Chaudhuri,
  Theriot, and Fletcher}}]{parekh2005}
\bibinfo{author}{\bibfnamefont{S.~H.} \bibnamefont{Parekh}},
  \bibinfo{author}{\bibfnamefont{O.}~\bibnamefont{Chaudhuri}},
  \bibinfo{author}{\bibfnamefont{J.~A.} \bibnamefont{Theriot}},
  \bibnamefont{and} \bibinfo{author}{\bibfnamefont{D.~A.}
  \bibnamefont{Fletcher}}, \bibinfo{journal}{Nature Cell Biology}
  \textbf{\bibinfo{volume}{7}}, \bibinfo{pages}{1219} (\bibinfo{year}{2005}).

\bibitem[{\citenamefont{Prass et~al.}(2006)\citenamefont{Prass, Jacobson,
  Mogilner, and Radmacher}}]{prass2006}
\bibinfo{author}{\bibfnamefont{M.}~\bibnamefont{Prass}},
  \bibinfo{author}{\bibfnamefont{K.}~\bibnamefont{Jacobson}},
  \bibinfo{author}{\bibfnamefont{A.}~\bibnamefont{Mogilner}}, \bibnamefont{and}
  \bibinfo{author}{\bibfnamefont{M.}~\bibnamefont{Radmacher}},
  \bibinfo{journal}{Journal of Cell Biology} \textbf{\bibinfo{volume}{174}},
  \bibinfo{pages}{767} (\bibinfo{year}{2006}).

\bibitem[{\citenamefont{Heinemann et~al.}(2011)\citenamefont{Heinemann,
  Doschke, and Radmacher}}]{heinemann2011}
\bibinfo{author}{\bibfnamefont{F.}~\bibnamefont{Heinemann}},
  \bibinfo{author}{\bibfnamefont{H.}~\bibnamefont{Doschke}}, \bibnamefont{and}
  \bibinfo{author}{\bibfnamefont{M.}~\bibnamefont{Radmacher}},
  \bibinfo{journal}{Biophysical Journal} \textbf{\bibinfo{volume}{100}},
  \bibinfo{pages}{1420} (\bibinfo{year}{2011}).

\bibitem[{\citenamefont{Zimmermann et~al.}(2012)\citenamefont{Zimmermann,
  Brunner, Enculescu, Goegler, Ehrlicher, K{\"a}s, and
  Falcke}}]{zimmermann2012}
\bibinfo{author}{\bibfnamefont{J.}~\bibnamefont{Zimmermann}},
  \bibinfo{author}{\bibfnamefont{C.}~\bibnamefont{Brunner}},
  \bibinfo{author}{\bibfnamefont{M.}~\bibnamefont{Enculescu}},
  \bibinfo{author}{\bibfnamefont{M.}~\bibnamefont{Goegler}},
  \bibinfo{author}{\bibfnamefont{A.}~\bibnamefont{Ehrlicher}},
  \bibinfo{author}{\bibfnamefont{J.}~\bibnamefont{K{\"a}s}}, \bibnamefont{and}
  \bibinfo{author}{\bibfnamefont{M.}~\bibnamefont{Falcke}},
  \bibinfo{journal}{Biophysical Journal} \textbf{\bibinfo{volume}{102}},
  \bibinfo{pages}{287} (\bibinfo{year}{2012}).

\bibitem[{\citenamefont{Maly and Borisy}(2001)}]{maly2001}
\bibinfo{author}{\bibfnamefont{I.~V.} \bibnamefont{Maly}} \bibnamefont{and}
  \bibinfo{author}{\bibfnamefont{G.~G.} \bibnamefont{Borisy}},
  \bibinfo{journal}{Proceedings of the National Academy of Sciences of the
  United States of America} \textbf{\bibinfo{volume}{98}},
  \bibinfo{pages}{11324} (\bibinfo{year}{2001}).

\bibitem[{\citenamefont{Schaub et~al.}(2007)\citenamefont{Schaub, Meister, and
  Verkhovsky}}]{schaub2007}
\bibinfo{author}{\bibfnamefont{S.}~\bibnamefont{Schaub}},
  \bibinfo{author}{\bibfnamefont{J.~J.} \bibnamefont{Meister}},
  \bibnamefont{and} \bibinfo{author}{\bibfnamefont{A.~B.}
  \bibnamefont{Verkhovsky}}, \bibinfo{journal}{Journal of Cell Science}
  \textbf{\bibinfo{volume}{120}}, \bibinfo{pages}{1491} (\bibinfo{year}{2007}).

\bibitem[{\citenamefont{Koestler et~al.}(2008)\citenamefont{Koestler, Auinger,
  Vinzenz, Rottner, and Small}}]{koestler2008}
\bibinfo{author}{\bibfnamefont{S.~A.} \bibnamefont{Koestler}},
  \bibinfo{author}{\bibfnamefont{S.}~\bibnamefont{Auinger}},
  \bibinfo{author}{\bibfnamefont{M.}~\bibnamefont{Vinzenz}},
  \bibinfo{author}{\bibfnamefont{K.}~\bibnamefont{Rottner}}, \bibnamefont{and}
  \bibinfo{author}{\bibfnamefont{J.~V.} \bibnamefont{Small}},
  \bibinfo{journal}{Nature Cell Biology} \textbf{\bibinfo{volume}{10}},
  \bibinfo{pages}{306} (\bibinfo{year}{2008}).

\bibitem[{\citenamefont{Weichsel et~al.}(2012)\citenamefont{Weichsel, Urban,
  Small, and Schwarz}}]{weichsel2012}
\bibinfo{author}{\bibfnamefont{J.}~\bibnamefont{Weichsel}},
  \bibinfo{author}{\bibfnamefont{E.}~\bibnamefont{Urban}},
  \bibinfo{author}{\bibfnamefont{J.~V.} \bibnamefont{Small}}, \bibnamefont{and}
  \bibinfo{author}{\bibfnamefont{U.~S.} \bibnamefont{Schwarz}},
  \bibinfo{journal}{Cytometry Part A} \textbf{\bibinfo{volume}{81A}},
  \bibinfo{pages}{496} (\bibinfo{year}{2012}).

\bibitem[{\citenamefont{Verkhovsky et~al.}(2003)\citenamefont{Verkhovsky,
  Chaga, Schaub, Svitkina, Meister, and Borisy}}]{verkhovsky2003}
\bibinfo{author}{\bibfnamefont{A.~B.} \bibnamefont{Verkhovsky}},
  \bibinfo{author}{\bibfnamefont{O.~Y.} \bibnamefont{Chaga}},
  \bibinfo{author}{\bibfnamefont{S.}~\bibnamefont{Schaub}},
  \bibinfo{author}{\bibfnamefont{T.~M.} \bibnamefont{Svitkina}},
  \bibinfo{author}{\bibfnamefont{J.~J.} \bibnamefont{Meister}},
  \bibnamefont{and} \bibinfo{author}{\bibfnamefont{G.~G.}
  \bibnamefont{Borisy}}, \bibinfo{journal}{Molecular Biology of the Cell}
  \textbf{\bibinfo{volume}{14}}, \bibinfo{pages}{4667} (\bibinfo{year}{2003}).

\bibitem[{\citenamefont{Carlsson}(2003)}]{carlsson2003}
\bibinfo{author}{\bibfnamefont{A.~E.} \bibnamefont{Carlsson}},
  \bibinfo{journal}{Biophysical Journal} \textbf{\bibinfo{volume}{84}},
  \bibinfo{pages}{2907} (\bibinfo{year}{2003}).

\bibitem[{\citenamefont{Schaus et~al.}(2007)\citenamefont{Schaus, Taylor, and
  Borisy}}]{schaus2007}
\bibinfo{author}{\bibfnamefont{T.~E.} \bibnamefont{Schaus}},
  \bibinfo{author}{\bibfnamefont{E.~W.} \bibnamefont{Taylor}},
  \bibnamefont{and} \bibinfo{author}{\bibfnamefont{G.~G.}
  \bibnamefont{Borisy}}, \bibinfo{journal}{Proceedings of the National Academy
  of Sciences of the United States of America} \textbf{\bibinfo{volume}{104}},
  \bibinfo{pages}{7086} (\bibinfo{year}{2007}).

\bibitem[{\citenamefont{Weichsel and Schwarz}(2010)}]{weichsel2010_2}
\bibinfo{author}{\bibfnamefont{J.}~\bibnamefont{Weichsel}} \bibnamefont{and}
  \bibinfo{author}{\bibfnamefont{U.~S.} \bibnamefont{Schwarz}},
  \bibinfo{journal}{Proceedings of the National Academy of Sciences of the
  United States of America} \textbf{\bibinfo{volume}{107}},
  \bibinfo{pages}{6304} (\bibinfo{year}{2010}).

\bibitem[{\citenamefont{Ti et~al.}(2011)\citenamefont{Ti, Jurgenson, Nolen, and
  Pollard}}]{ti2011}
\bibinfo{author}{\bibfnamefont{S.-C.} \bibnamefont{Ti}},
  \bibinfo{author}{\bibfnamefont{C.~T.} \bibnamefont{Jurgenson}},
  \bibinfo{author}{\bibfnamefont{B.~J.} \bibnamefont{Nolen}}, \bibnamefont{and}
  \bibinfo{author}{\bibfnamefont{T.~D.} \bibnamefont{Pollard}},
  \bibinfo{journal}{Proceedings of the National Academy of Sciences of the
  United States of America} \textbf{\bibinfo{volume}{108}},
  \bibinfo{pages}{E463} (\bibinfo{year}{2011}).

\bibitem[{\citenamefont{Mullins et~al.}(1998)\citenamefont{Mullins, Heuser, and
  Pollard}}]{mullins1998}
\bibinfo{author}{\bibfnamefont{R.~D.} \bibnamefont{Mullins}},
  \bibinfo{author}{\bibfnamefont{J.~A.} \bibnamefont{Heuser}},
  \bibnamefont{and} \bibinfo{author}{\bibfnamefont{T.~D.}
  \bibnamefont{Pollard}}, \bibinfo{journal}{Proceedings of the National Academy
  of Sciences of the United States of America} \textbf{\bibinfo{volume}{95}},
  \bibinfo{pages}{6181} (\bibinfo{year}{1998}).

\bibitem[{\citenamefont{Blanchoin et~al.}(2000)\citenamefont{Blanchoin, Amann,
  Higgs, Marchand, Kaiser, and Pollard}}]{blanchoin2000}
\bibinfo{author}{\bibfnamefont{L.}~\bibnamefont{Blanchoin}},
  \bibinfo{author}{\bibfnamefont{K.~J.} \bibnamefont{Amann}},
  \bibinfo{author}{\bibfnamefont{H.~N.} \bibnamefont{Higgs}},
  \bibinfo{author}{\bibfnamefont{J.~B.} \bibnamefont{Marchand}},
  \bibinfo{author}{\bibfnamefont{D.~A.} \bibnamefont{Kaiser}},
  \bibnamefont{and} \bibinfo{author}{\bibfnamefont{T.~D.}
  \bibnamefont{Pollard}}, \bibinfo{journal}{Nature}
  \textbf{\bibinfo{volume}{404}}, \bibinfo{pages}{1007} (\bibinfo{year}{2000}).

\bibitem[{\citenamefont{Svitkina and Borisy}(1999)}]{svitkina1999}
\bibinfo{author}{\bibfnamefont{T.~M.} \bibnamefont{Svitkina}} \bibnamefont{and}
  \bibinfo{author}{\bibfnamefont{G.~G.} \bibnamefont{Borisy}},
  \bibinfo{journal}{The Journal of Cell Biology}
  \textbf{\bibinfo{volume}{145}}, \bibinfo{pages}{1009} (\bibinfo{year}{1999}).

\bibitem[{\citenamefont{Vinzenz et~al.}(2012)\citenamefont{Vinzenz, Nemethova,
  Schur, Mueller, Narita, Urban, Winkler, Schmeiser, Koestler, Rottner
  et~al.}}]{vinzenz2012}
\bibinfo{author}{\bibfnamefont{M.}~\bibnamefont{Vinzenz}},
  \bibinfo{author}{\bibfnamefont{M.}~\bibnamefont{Nemethova}},
  \bibinfo{author}{\bibfnamefont{F.}~\bibnamefont{Schur}},
  \bibinfo{author}{\bibfnamefont{J.}~\bibnamefont{Mueller}},
  \bibinfo{author}{\bibfnamefont{A.}~\bibnamefont{Narita}},
  \bibinfo{author}{\bibfnamefont{E.}~\bibnamefont{Urban}},
  \bibinfo{author}{\bibfnamefont{C.}~\bibnamefont{Winkler}},
  \bibinfo{author}{\bibfnamefont{C.}~\bibnamefont{Schmeiser}},
  \bibinfo{author}{\bibfnamefont{S.}~\bibnamefont{Koestler}},
  \bibinfo{author}{\bibfnamefont{K.}~\bibnamefont{Rottner}},
  \bibnamefont{et~al.}, \bibinfo{journal}{Journal of Cell Science}
  \textbf{\bibinfo{volume}{125}}, \bibinfo{pages}{2775} (\bibinfo{year}{2012}).

\bibitem[{\citenamefont{Chaudhuri et~al.}(2009)\citenamefont{Chaudhuri, Parekh,
  Lam, and Fletcher}}]{chaudhuri2009}
\bibinfo{author}{\bibfnamefont{O.}~\bibnamefont{Chaudhuri}},
  \bibinfo{author}{\bibfnamefont{S.~H.} \bibnamefont{Parekh}},
  \bibinfo{author}{\bibfnamefont{W.~A.} \bibnamefont{Lam}}, \bibnamefont{and}
  \bibinfo{author}{\bibfnamefont{D.~A.} \bibnamefont{Fletcher}},
  \bibinfo{journal}{Nature Methods} \textbf{\bibinfo{volume}{6}},
  \bibinfo{pages}{383} (\bibinfo{year}{2009}).

\bibitem[{\citenamefont{Millius et~al.}(2012)\citenamefont{Millius, Watanabe,
  and Weiner}}]{millius2012}
\bibinfo{author}{\bibfnamefont{A.}~\bibnamefont{Millius}},
  \bibinfo{author}{\bibfnamefont{N.}~\bibnamefont{Watanabe}}, \bibnamefont{and}
  \bibinfo{author}{\bibfnamefont{O.~D.} \bibnamefont{Weiner}},
  \bibinfo{journal}{Journal of Cell Science} \textbf{\bibinfo{volume}{125}},
  \bibinfo{pages}{1165} (\bibinfo{year}{2012}).

\bibitem[{\citenamefont{si}(2012)\citenamefont{si}}]{si}
  \bibinfo{journal}{See supplemental material for more details.}

\end{thebibliography}


\end{document}